\documentstyle[12pt,a4,psfig,axodraw]{article}
\begin{document}
\baselineskip 18pt
\def\today{\ifcase\month\or
 January\or February\or March\or April\or May\or June\or
 July\or August\or September\or October\or November\or December\fi
 \space\number\day, \number\year}
\def\thebibliography#1{\section*{References\markboth
 {References}{References}}\list
 {[\arabic{enumi}]}{\settowidth\labelwidth{[#1]}
 \leftmargin\labelwidth
 \advance\leftmargin\labelsep
 \usecounter{enumi}}
 \def\newblock{\hskip .11em plus .33em minus .07em}
 \sloppy
 \sfcode`\.=1000\relax}
\let\endthebibliography=\endlist
\def\MPLA#1#2#3{Mod. Phys. Lett. {\bf A#1} (#2) #3}
\def\PRD#1#2#3{Phys. Rev. {\bf D#1} (#2) #3}
\def\NPB#1#2#3{Nucl. Phys. {\bf B#1} (#2) #3}
\def\PTP#1#2#3{Prog. Theor. Phys. {\bf #1} (#2) #3}
\def\ZPC#1#2#3{Z. Phys. {\bf C#1} (#2) #3}
\def\EPJC#1#2#3{Eur. Phys. J. {\bf C#1} (#2) #3}
\def\PLB#1#2#3{Phys. Lett. {\bf B#1} (#2) #3}
\def\PRL#1#2#3{Phys. Rev. Lett. {\bf #1} (#2) #3}
\def\PRep#1#2#3{Phys. Rep. {\bf #1} (#2) #3}
\def\RMP#1#2#3{Rev. Mod. Phys. {\bf #1} (#2) #3}
\def\gsim{~{\rlap{\lower 3.5pt\hbox{$\mathchar\sim$}}\raise 1pt\hbox{$>$}}\,}
\def\lsim{~{\rlap{\lower 3.5pt\hbox{$\mathchar\sim$}}\raise 1pt\hbox{$<$}}\,}
\def\ov{\overline}
\def\wt{\widetilde}
\def\r2{\sqrt 2}
\def\half{\frac{1}{2}}
\def\bsg{b\rightarrow s\gamma}
\def\bsgg{b\rightarrow s\gamma g}
\def\BSgamma{B\rightarrow X_s\gamma}
\def\BsBs{B_s\ov{B_s}}
\def\mqu#1{m_{u#1}}
\def\mqd#1{m_{d#1}}
\def\VT{V_{32}^*V_{33}}
\def\VU{V_{42}^*V_{43}}
\def\ZBS{(V^\dagger V)_{23}}
\newcommand{\beq}{\begin{equation}}
\newcommand{\eeq}{\end{equation}}
\newcommand{\bea}{\begin{eqnarray}}
\newcommand{\eea}{\end{eqnarray}}
\begin{titlepage}
\begin{flushright}
\begin{tabular}{l}
{OCHA-PP-170} \\
{KEK-TH-747}\\
\end{tabular}
\end{flushright}
\vskip 0.5 true cm 
\begin{center}
{\large {\bf Large effects on $\BsBs$ mixing by vector-like quarks  \\}}  
\vskip 2.0 true cm
\renewcommand{\thefootnote}
{\fnsymbol{footnote}}
Mayumi Aoki$^1$\footnote{JSPS Fellow.}, 
Gi-Chol Cho$^2$, Makiko Nagashima$^3$, 
and Noriyuki Oshimo$^2$ 
\\
\vskip 0.5 true cm 
{\it $^1$Theory Group, KEK, Tsukuba, Ibaraki 305-0801, Japan}  \\
{\it $^2$Department of Physics}  \\
{\it Ochanomizu University, Bunkyo-ku, Tokyo 112-8610, Japan}  \\
{\it $^3$Graduate School of Humanities and Sciences}  \\
{\it Ochanomizu University, Bunkyo-ku, Tokyo 112-8610, Japan}  \\
\end{center}

\vskip 3.0 true cm

\centerline{\bf Abstract}
\medskip
     We calculate the contributions of the vector-like quark model 
to $\BsBs$ mixing, taking into account the 
constraints from the decay $B\to X_s\gamma$.   
In this model the neutral bosons mediate flavor-changing interactions  
at the tree level.  
However, $\BsBs$ mixing is dominated by 
contributions from the box diagrams with the top quark and 
the extra up-type quark.  
In sizable ranges of the model parameters, the mixing parameter $x_s$ 
is much different from the standard model prediction.  

\vspace{3mm}
\noindent
PACS number(s): 
12.15.Ff, 12.60.-i, 14.40.Nd
\medskip

\end{titlepage}

\newpage 
\setcounter{footnote}{0}
 
     The standard model (SM) may have to be extended to 
describe physics around or above the electroweak scale.
Various works have thus discussed phenomena involving    
the $B$ meson which are sensitive to new physics \cite{review}.   
For instance, the radiative $B$-meson decay $B\to X_s\gamma$ 
and $B\ov{B}$ mixing could receive non-trivial contributions 
from supersymmetry \cite{oshimo, branco}.  
The vector-like quark model (VQM) could also affect  
these processes of flavor-changing neutral 
current \cite{nir, aoki}.  

     In this letter we discuss $\BsBs$ mixing within the 
framework of the VQM which is one of the minimal extensions of the SM.  
The mixing parameter $x_s$ 
is evaluated and its dependencies on the model parameters are analyzed.  
These model parameters are constrained by the branching ratio of 
the decay $B\to X_s\gamma$.  
Even under these constraints, the value of $x_s$ can be much different 
from the SM prediction.  

     The VQM incorporates extra quarks whose left-handed 
components, as well as right-handed ones, are singlets under 
the SU(2) gauge transformation.  
Then, the interactions of quarks 
become different from those in the SM.   
The Cabibbo-Kobayashi-Maskawa (CKM) matrix for the interactions 
with the $W$ boson is extended and not unitary.  
The $Z$ boson couples directly to the quarks with different flavors.  
The neutral Higgs boson also mediates 
flavor-changing interactions at the tree level.  

     Our study concentrates on 
$\BsBs$ mixing at the one-loop level through box diagrams.  
The contributions at the tree level have already been analyzed 
in the literature.  
The order of these tree-level diagrams is lower than the 
box diagrams of the SM.  
However, the experimental bounds on the branching ratio of 
the decay $B\to K\mu^+\mu^-$ suggest very 
weak couplings for the flavor-changing interactions at the tree level.   
Consequently, the tree-level contributions turn out to be smaller than 
the one-loop contributions.  
On the other hand, the new box diagrams in the VQM are 
expected to contribute at the same order as those of the SM.  
Although these new contributions do not yield a drastic change 
from the SM prediction, precise measurements in near-future 
experiments, such as BTeV and LHCb, may be able to distinguish 
the VQM and the SM.  
Indeed, the VQM could give sizable new contributions to 
the decay $B\to X_s\gamma$ at the one-loop level \cite{aoki}, 
imposing non-trivial constraints on the model.     

     We assume, for definiteness, that there exist two vector-like quarks 
$U$ and $D$ with electric charge $2/3$ and $-1/3$, respectively.  
The CKM matrix $V$ is then enlarged to be a $4\times 4$ matrix and 
expressed as 
\begin{equation}
      V_{ab} = \sum_{i=1}^3(A_L^{u\dagger})_{ai}(A_L^d)_{ib},  
\end{equation}
$A^u_L$ and $A^d_L$ being unitary matrices which diagonalize 
the mass matrices for up-type and down-type quarks, respectively.   
It should be noted that the matrix $V$ is not unitary:   
\beq
  (V^\dagger V)_{ab}=\delta_{ab}-A^{d*}_{L4a}A^d_{L4b}.  
\eeq
The interaction Lagrangian for the quarks with the $W$ and 
Goldstone bosons is given by 
\begin{eqnarray}
 {\cal L} &=& \frac{g}{\r2}\sum_{a,b=1}^4\overline{u^a}V_{ab}\gamma^\mu               
                   \frac{1-\gamma_5}{2}d^bW_\mu^\dagger   \nonumber  \\ 
 & & +\frac{g}{\r2}\sum_{a,b=1}^4\overline{u^a}V_{ab}\left\{\frac{m_{ua}}{M_W}            
                   \left(\frac{1-\gamma_5}{2}\right)
              - \frac{m_{db}}{M_W}\left(\frac{1+\gamma_5}{2}\right)\right\}
                d^bG^\dagger  \nonumber \\
         & & +{\rm h.c.}.    
\label{Wlagrangian}
\end{eqnarray}
Here, the mass eigenstates for the up-type and down-type quarks 
are denoted by $u^a$ and $d^b$, 
$a$ and $b$ being the generation indices, 
and $m_{ua}$ and $m_{db}$ represent the corresponding mass eigenvalues.  
These eigenstates will also be called as $(u,c,t,U)$ and $(d,s,b,D)$.  
The interaction Lagrangian for the down-type quarks 
with the $Z$, Higgs, and Goldstone bosons is given by  
\begin{eqnarray}
{\cal L} &=& -\frac{g}{\cos\theta_W}\sum_{a,b=1}^4
 \overline{d^a}\gamma^\mu
            \left\{-\half\left(V^\dagger V\right)_{ab}\frac{1-\gamma_5}{2}
    +\frac{1}{3}\sin^2\theta_W\delta_{ab}\right\}d^bZ_\mu   
\nonumber  \\
   & & -\frac{g}{2}\sum_{a,b=1}^4\overline{d^a}\left(V^\dagger V\right)_{ab}
\left\{\frac{m_{da}}{M_W}\left(\frac{1-\gamma_5}{2}\right)
    +\frac{m_{db}}{M_W}\left(\frac{1+\gamma_5}{2}\right)\right\}d^bH^0   
\nonumber  \\
     & & +i\frac{g}{2}\sum_{a,b=1}^4\overline{d^a}\left(V^\dagger V\right)_{ab}
\left\{\frac{m_{da}}{M_W}\left(\frac{1-\gamma_5}{2}\right)
  -\frac{m_{db}}{M_W}\left(\frac{1+\gamma_5}{2}\right)\right\}d^bG^0.  
\label{Zlagrangian} 
\end{eqnarray}
Since the matrix $V$ is not unitary, there appear flavor-changing 
interactions at the tree level.     
The Lagrangians in Eqs. (\ref{Wlagrangian}) and (\ref{Zlagrangian}) 
contain also new sources of $CP$ violation \cite{asakawa}.    

     The amount of $\BsBs$ mixing is described by 
the mixing parameter $x_s$, which becomes  
\bea
x_s &=& \frac{G_F}{3\sqrt 2} (\sqrt {B_{B_s}} f_{B_s})^2
               M_{B_s} \eta_{QCD} \tau_{B_s} \nonumber \\
          & & \left |\{\ZBS\}^2 + \frac{G_F}{\sqrt 2\pi^2} M_W^2
              \sum_{a,b=1}^4 V^{\ast}_{a2}V_{a3} V^{\ast}_{b2}V_{b3}
              S(r_a , r_b) 
              \right |,             \label{xs}  \\
S(r_a,r_b) &=& \frac{4-7r_ar_b}{4(1-r_a)(1-r_b)} \nonumber \\ 
                 &+& \frac{4-8r_b+r_ar_b}{4(1-r_a)^2(r_a-r_b)}r_a^2\ln r_a
                      + \frac{4-8r_a+r_ar_b}{4(1-r_b)^2(r_b-r_a)}r_b^2\ln r_b, 
\eea
with $r_a = m_{ua}^2/M_W^2$.  
The first term in Eq. (\ref{xs}) comes from the diagram 
exchanging the $Z$ boson at the tree level.  
The contribution from the tree-level diagram exchanging the 
Higgs boson is suppressed by a factor of $(m_b/M_W)^2$ 
and thus negligible.  
The second term arises from the box diagrams.  
Neglecting the masses of the $u$ and $c$ quarks, we obtain   
\bea
 & & \sum_{a,b=1}^4 V^{\ast}_{a2}V_{a3} V^{\ast}_{b2}V_{b3} S(r_a , r_b) 
  = \{(V^{\dagger}V)_{23}\}^2   
 +  2\sum_{a=3,4}(V^{\dagger}V)_{23}V^{\ast}_{a2}V_{a3} I_1(r_a) \nonumber \\
 & &  ~\ ~\ ~\ ~\ + \sum_{a=3,4}(V^{\ast}_{a2}V_{a3})^2 I_2(r_a)
    +  2V^{\ast}_{32}V_{33}V^{\ast}_{42}V_{43} I_3(r_3,r_4), 
\eea
where the functions $I_1$, $I_2$, $I_3$ are defined by 
\bea
 I_1(r_a) &=& S(0,r_a)-S(0,0),  \\
 I_2(r_a) &=& S(r_a,r_a)-2S(0,r_a)+S(0,0),  \\
 I_3(r_3,r_4) &=& S(r_3,r_4)-S(0,r_3)-S(0,r_4)+S(0,0). 
\eea 
In our numerical calculations we take 
the bag factor $B_{B_s}$, the decay constant $f_{B_s}$, 
the meson mass $M_{B_s}$, and the meson life time $\tau_{B_s}$ 
for $\sqrt {B_{B_s}}f_{B_s}=267$ MeV \cite{lattice}, $M_{B_s}=5.37$ GeV, and 
$\tau_{B_s}=1.49$ ps \cite{pdg}.  
The QCD correction factor $\eta_{QCD}$ is set 
at 0.55 \cite{buras}.  

     The mixing parameter $x_s$ depends on the $U$-quark mass  
$m_U$ and the CKM matrix elements $\VT$, $\VU$, $\ZBS$.  
The mass should be larger than the top-quark mass.  
The matrix elements are related to $V_{12}$, $V_{13}$, $V_{22}$, 
and $V_{23}$ which have been directly measured in experiments.  
Their experimental values \cite{pdg} give a constraint 
\beq
0.03<|\VT+\VU-\ZBS|<0.05.  
\label{constraint1}
\eeq
As seen in Eq. (\ref{Zlagrangian}),
the value of $\ZBS$ determines the flavor-changing interactions at the 
tree level.  
From the upper bounds on the branching ratio of the decay 
$B\rightarrow K\mu^+\mu^-$ \cite{pdg}, a constraint 
\beq
 |\ZBS| <2.0\times 10^{-3}
\label{constraint2}
\eeq
is derived \footnote{The present experimental bounds do not give 
non-trivial constraints on $\VT$ and $\VU$ which affect the decay 
through one-loop diagrams.  
For further discussions, see Ref. \cite{ahmady}.}.  
The $U$-quark mass and the CKM matrix 
elements are related to each other through the mass 
matrices of the up-type and down-type quarks.  
However, there are many unknown factors for 
these mass matrices, leaving open various possibilities 
for the relations \cite{higuchi}.  
Therefore, for our numerical analyses, we assume that 
the model parameters $m_U$, $\VT$, $\VU$, and $\ZBS$ are 
independent of each other.  
For simplicity, these matrix elements are taken 
as real.  

\begin{figure}
\psfig{file=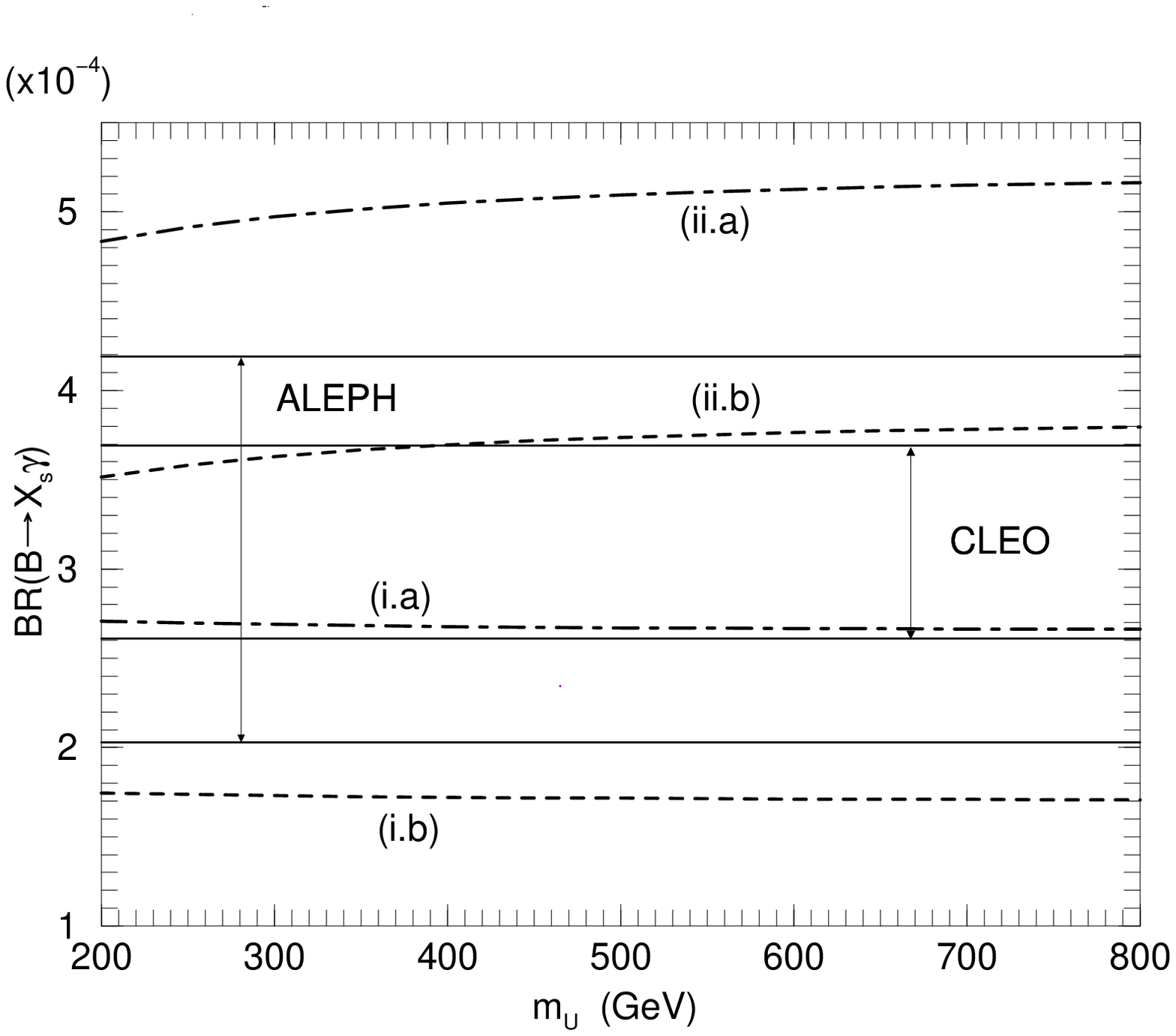,width=14cm,height=12cm,angle=0}
\caption{
The branching ratio of $B\to X_s\gamma$ for $\VT=0.035$ 
as a function of the $U$-quark mass.  Four curves correspond to 
the four parameter sets given in Table \ref{tab35}.  
The 1-$\sigma$ allowed range of the branching ratio from
CLEO \cite{cleo} and ALEPH \cite{aleph} are also shown.
} 
\label{figbs35}
\end{figure}

\begin{figure}
\psfig{file=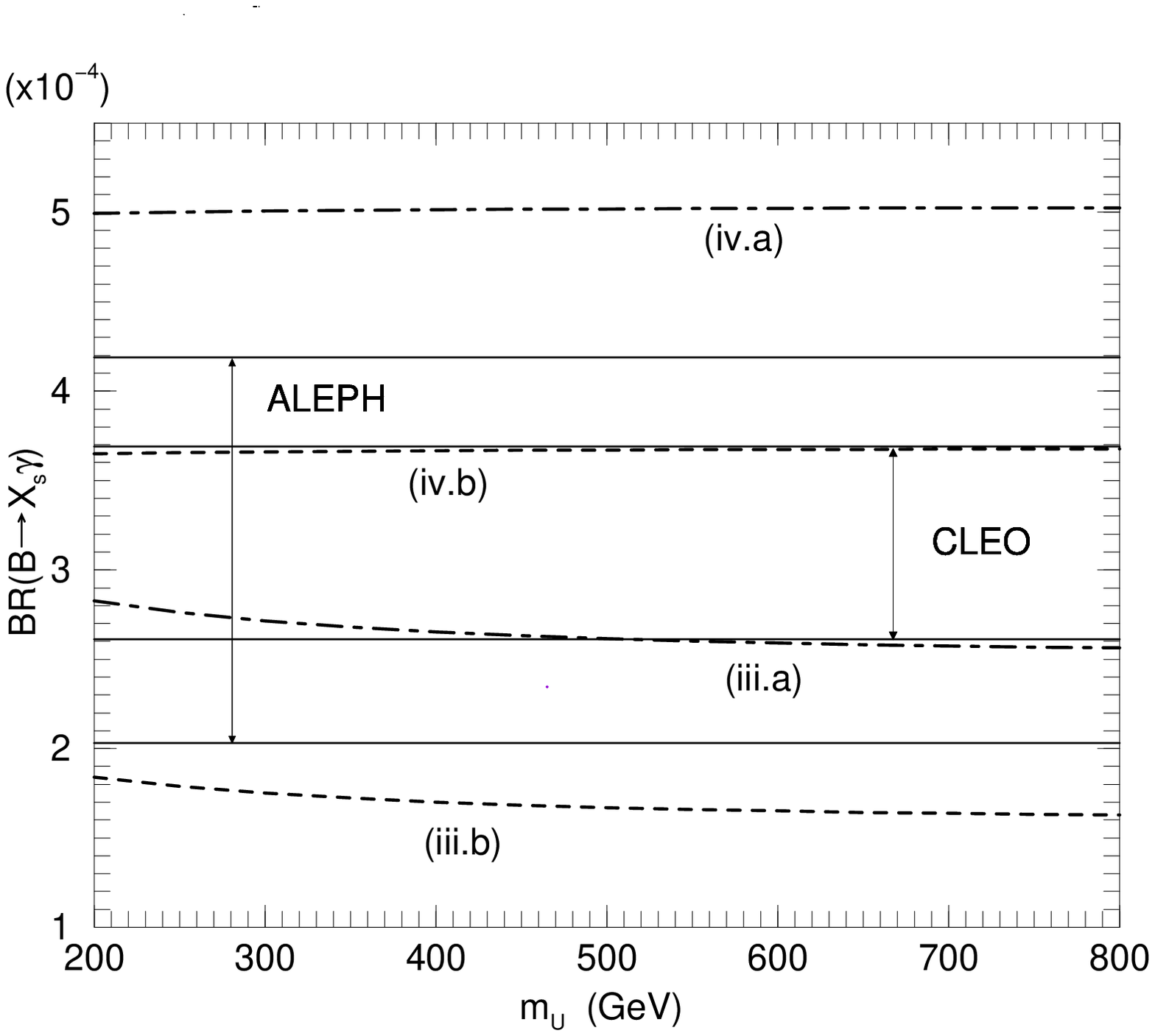,width=14cm,height=12cm,angle=0}
\caption{
The branching ratio of $B\to X_s\gamma$ for $\VT=0.045$ 
as a function of the $U$-quark mass.  Four curves correspond to 
the four parameter sets given in Table \ref{tab45}.
The 1-$\sigma$ allowed range of the branching ratio from
CLEO \cite{cleo} and ALEPH \cite{aleph} are also shown.  
} 
\label{figbs45}
\end{figure}

     The model parameters are further constrained \cite{aoki} from 
the branching  ratio of $B\to X_s\gamma$, which has been measured 
by CLEO \cite{cleo} and ALEPH \cite{aleph} as, respectively, 
$Br(B\to X_s\gamma)=(3.15\pm 0.35\pm 0.32\pm 0.26)\times 10^{-4}$ 
(1-$\sigma$) and  $Br(B\to X_s\gamma)=(3.11\pm 0.80\pm 0.72)\times 10^{-4}$ 
(1-$\sigma$).  
We show its predicted branching ratio in Figs. \ref{figbs35} 
and \ref{figbs45} for $\VT=0.035$ and $\VT=0.045$, respectively, 
as a function of the $U$-quark mass.  
The values of $\VU$ and $\ZBS$ are listed in Tables \ref{tab35} 
and \ref{tab45}.  
The experimental bounds are shown by solid lines.  
For $\VT=0.035$, the ranges $\VU<-0.002$ and $0.010<\VU$ are 
not allowed from the CLEO bounds, irrespectively of 
the value for $\ZBS$.  
For $\VT=0.045$, similarly excluded are the ranges $\VU<-0.011$ 
and $0.001<\VU$.  
The allowed ranges of $\VU$ and $\ZBS$ 
do not depend very much 
on the $U$-quark mass.  
A rather large part of  the region which satisfies Eqs. (\ref{constraint1}) 
and (\ref{constraint2}) becomes inconsistent with $B\to X_s\gamma$.  
      
\begin{figure}
\psfig{file=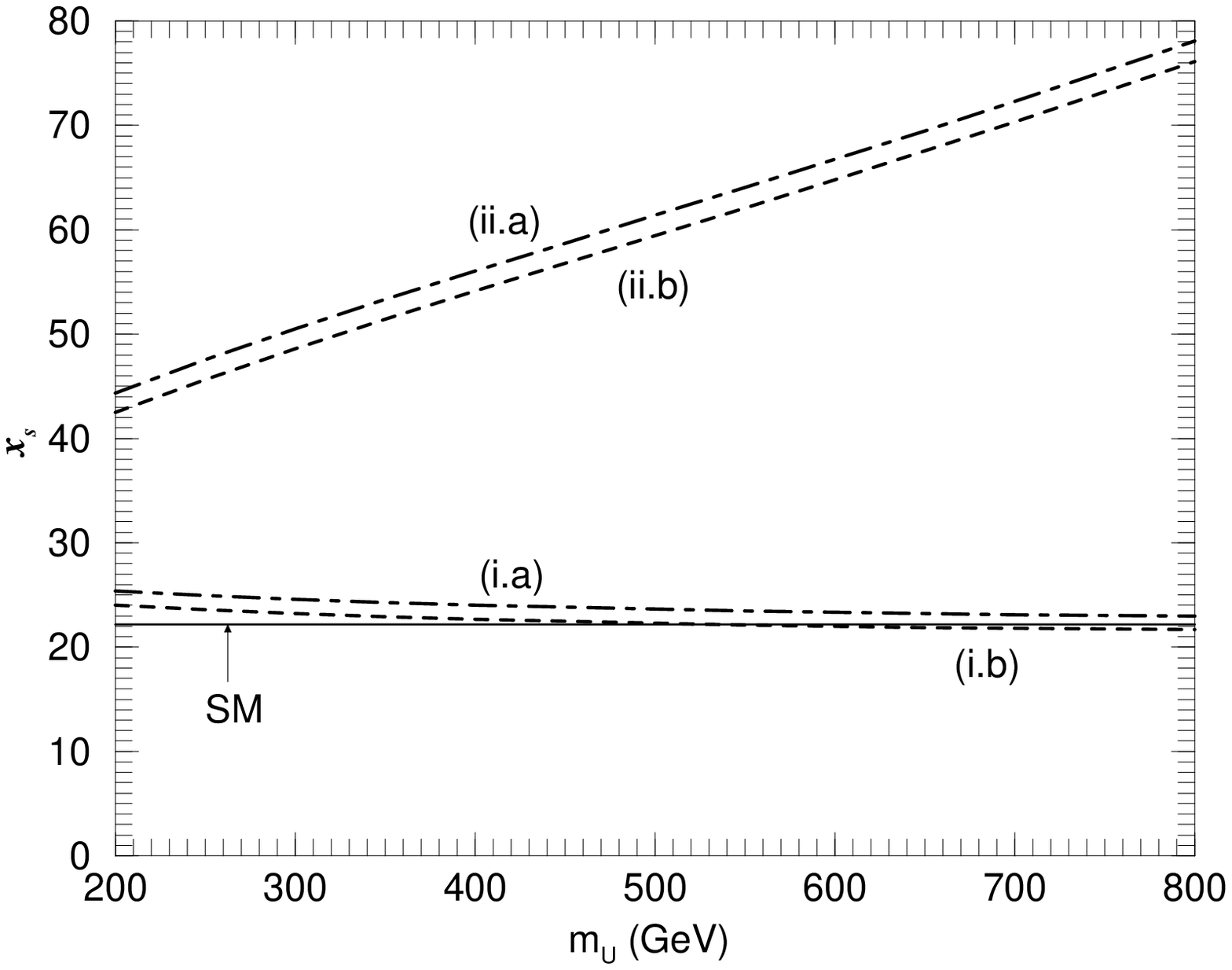,width=14cm,height=12cm,angle=0}
\caption{
The mixing parameter $x_s$ for $\VT=0.035$ 
as a function of the $U$-quark mass.  
Four curves correspond to 
the four parameter sets given in Table \ref{tab35}.  
The solid horizontal line represents the SM prediction for $m_t= 177$GeV .
} 
\label{figbb35}
\end{figure}

\begin{figure}
\psfig{file=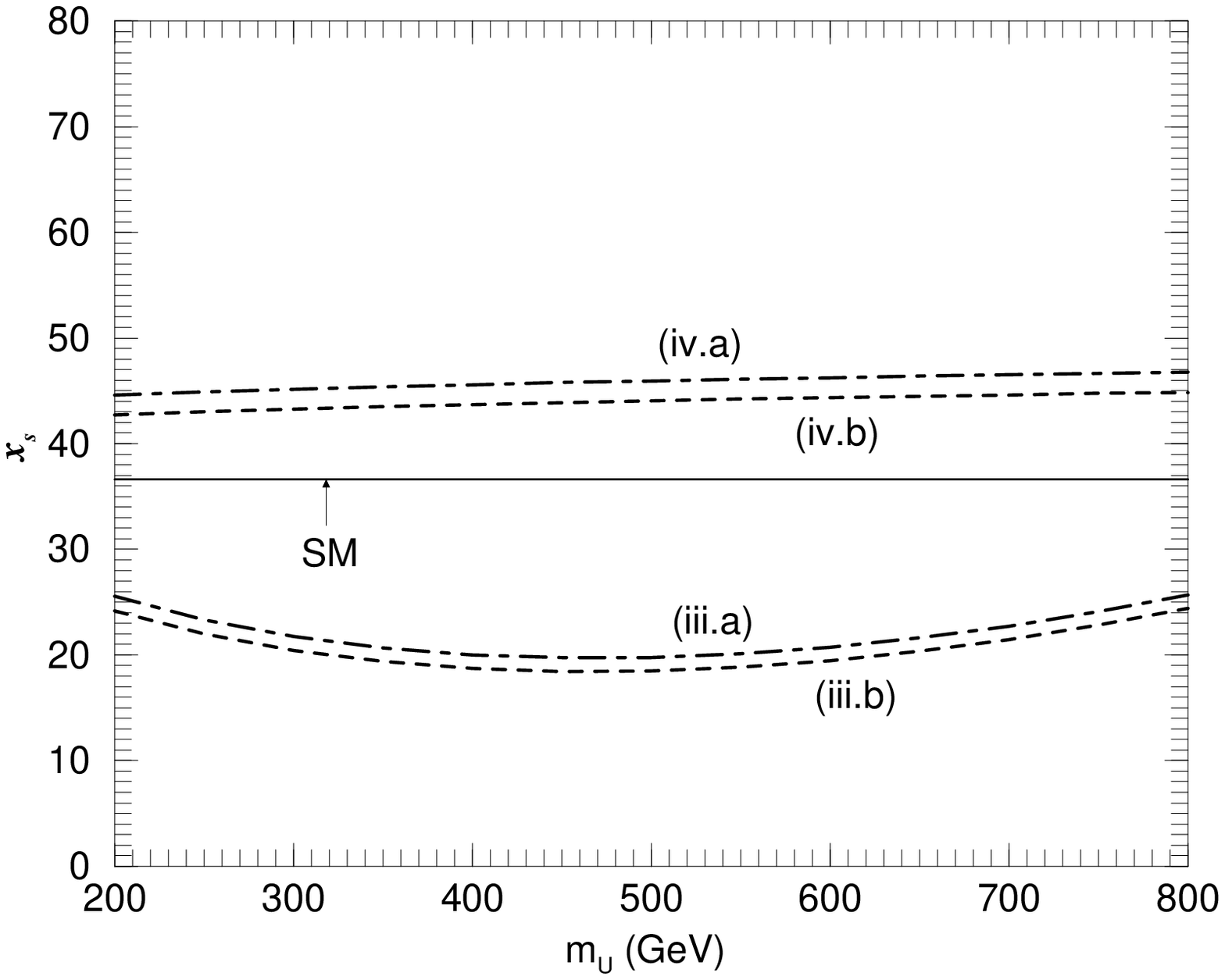,width=14cm,height=12cm,angle=0}
\caption{
The mixing parameter $x_s$ for $\VT=0.045$ 
as a function of the $U$-quark mass.  
Four curves correspond to 
the four parameter sets given in Table \ref{tab45}.  
The solid horizontal line represents the SM prediction for $m_t= 177$GeV .
} 
\label{figbb45}
\end{figure}

     We now evaluate $\BsBs$ mixing within the obtained 
parameter ranges consistent with the experiments.  
In Figs. \ref{figbb35} and \ref{figbb45} the mixing parameter $x_s$ 
is shown for the same parameter sets as those 
in Figs. \ref{figbs35} and \ref{figbs45}, respectively.  
The solid line represents the SM prediction.  
The parameter $x_s$ has a value between the curves (i) and (ii) 
in Fig. \ref{figbb35} and the curves (iii) and (iv) in Fig. \ref{figbb45} 
for the ranges of $\VU$ allowed by the CLEO bounds.  
A larger value for $\VU$ leads to a larger value for $x_s$.  
For $\VT=0.035$ and $-0.002\leq\VU\leq 0.010$, 
the value of $x_s$ is larger than the SM value, 
and their difference can be as much as a factor of two or more.  
As the value of $\VT$ increases, this difference becomes small.  
Still, a deviation by a factor of 0.5 of the SM value can 
occur for $\VT=0.045$ and $-0.011\leq\VU\leq 0.001$.  
The value of $\ZBS$ does not affect much $x_s$, while 
a larger mass of the $U$ quark tends to give a larger value for $x_s$.  
The magnitude of the tree-level contribution to $x_s$  
is at most 20 percent of the SM value.  
It should be noted that the value of $x_s$ may be 
smaller than the SM value, which stands in contrast to the 
prediction by the supersymmetric standard model \cite{branco}.  

     In the SM, the measurement of $x_s$ determines the value of $\VT$, 
which is examined from the point of unitarity for the CKM matrix.  
Within the present accuracies for the values of $V^*_{12}V_{13}$ and 
$V^*_{22}V_{23}$, the range $x_s>50$ leads to unitarity violation.  
More precise measurements of $V^*_{12}V_{13}$ and $V^*_{22}V_{23}$, 
together with smaller errors in calculating $B_{B_s}$ and $f_{B_s}$, 
will make it possible to examine the range $x_s<50$.  
Alternatively, the value of $\VT$ may be compared with that 
measured by the top-quark decays.   
A possible contradiction in these examinations within the framework 
of the SM could imply the existence of the vector-like quarks.  
  
     In summary, we have studied the effects of the VQM 
on $\BsBs$ mixing.  
The contribution of the $W$-mediated diagrams at the one-loop 
level could be sizably different from that in the SM.  
On the other hand, the $Z$-mediated diagrams at the tree level 
cause merely a small effect.  
The VQM is constrained from experimental results 
for 
the radiative decay $B\to X_s\gamma$.  
Under these constraints, the mixing parameter $x_s$ can 
non-trivially be larger or smaller than the SM prediction.  
Its value could be more than twice the SM value.  
The measurement of $x_s$ provides a useful test for the VQM.  

\smallskip 
        
     The authors thank T. Onogi and M. Yamauchi for discussions.  
The work of M.A. is supported in part by a Grant-in-Aid for 
Scientific Research from the Ministry of Education, Science and
Culture, Japan.

\newpage 
\begin{table}
\begin{center}
\begin{tabular}{c r r r r}
\hline
   & (i.a) & (i.b) & (ii.a) & (ii.b) \\
\hline
 $\VU$   & $-0.002$ & $-0.002$ & 0.010 & 0.010 \\
 $\ZBS$  & $-0.002$ & 0.002 & $-0.002$ & 0.002 \\
\hline
\end{tabular}
\end{center}
\caption{The values of $\VU$ and $\ZBS$  
in Figs. \ref{figbs35} and \ref{figbb35}.}
\label{tab35}
\end{table}
\begin{table}
\begin{center}
\begin{tabular}{c r r r r}
\hline
   & (iii.a) & (iii.b) & (iv.a) & (iv.b) \\
\hline
 $\VU$   & $-0.011$ & $-0.011$ & 0.001 & 0.001 \\
 $\ZBS$  & $-0.002$ & 0.002 & $-0.002$ & 0.002 \\
\hline
\end{tabular}
\end{center}
\caption{The values of $\VU$ and $\ZBS$  
in Figs. \ref{figbs45} and \ref{figbb45}.}
\label{tab45}
\end{table}

\pagebreak

\pagebreak

\pagebreak

\pagebreak

\end{document}